\documentclass[a4paper,11pt]{article}
\usepackage{jheppub} 
\usepackage{lineno}
\usepackage{amsmath}

\def\bc{\begin{center}}

\def\ec{\end{center}}
\def\be{\begin{eqnarray}}
\def\ee{\end{eqnarray}}

\def\blue{\color{blue}}

\title{\boldmath Revisit the entanglement entropy with gravitational anomaly}

\author[a]{Peng-Zhang He}
\author[a,b]{, Hai-Qing Zhang}
\affiliation[a]{Center for Gravitational Physics, Department of Space Science, Beihang University, Beijing 100191, China}
\affiliation[b]{Peng Huanwu Collaborative Center for Research and Education, Beihang University, Beijing 100191, China}

\emailAdd{hepzh@buaa.edu.cn}
\emailAdd{hqzhang@buaa.edu.cn}

\abstract{In this paper we study the entanglement entropy in the CFT$_2$, whose gravity dual is AdS$_3$ spacetime  with a Chern-Simons term. Using the generalized Rindler method, we obtain the Rindler transformation in the two-dimensional planar CFT and compute the entanglement entropy of the CFT with gravitational anomalies. The conditions under which the entanglement entropy may have anomalous contributions is also discussed. In addition, we present a relatively general form of the Rindler AdS metric and compute its thermal entropy, which agrees with the entanglement entropy in the field theory. Moreover, we utilize the conformal transformation, which maps a cylinder to a plane, to compute the entanglement entropy of the CFT residing on a cylinder, as well as the entanglement entropy of the CFT at finite temperature on a plane.  The corresponding contribution of the Chern-Simons term in gravity to the black hole thermal entropy is also obtained from this approach. These results are important for further understandings of the two-dimensional CFT with gravitational anomalies.}

\begin{document}
\maketitle
\flushbottom

	\section{Introduction}
	Quantum entanglement is a quantum phenomenon that has no classical counterpart~\cite{RevModPhys.81.865}. Entanglement entropy is a measure of quantum entanglement that reflects how entanglement is stored in quantum states. For instance,  the Hilbert space of a biparticle system can be written as the product of two factors
	\begin{equation}
	\mathcal{H} =\mathcal{H} _A\otimes \mathcal{H} _{\bar{A}},
	\end{equation} 
where $\mathcal{H} _A$ and $\mathcal{H}_{\bar{A}}$ are the Hilbert space of the two subsystems $A$ and $\bar A$, respectively.  The von Neumann entropy $S_A$ of the subsystem $A$ is defined as
\begin{equation}
S_A=-\mathrm{Tr}\rho _A\log \rho _A,\qquad \rho _A=\mathrm{Tr}_{\bar{A}}\rho 
\end{equation}
where $\rho$ is the density matrix of the total system. $S_A$ is also called the entanglement entropy~\cite{hartman2015lectures}. In a continuous QFT, if the space is divided into two parts $A$ and $B$, with the boundary $\partial A$, the computation of the entanglement entropy is not easy since the result is divergent~\cite{Calabrese:2004eu,Calabrese:2009qy}. A UV cutoff must be introduced to obtain a finite result.  Entanglement entropy in the two-dimensional conformal field theory (CFT) is an active area of research, in which the entanglement entropy can be systematically calculated in many cases~\cite{Calabrese:2004eu}, such as:
 \begin{itemize}
 	\item A CFT sits on an infinite plane, and consider an interval $A$ of length $\ell$ on this plane. Its entanglement entropy is
 	\begin{equation}
 		S_{A} \sim(c / 3) \log (\ell / a),
 	\end{equation}
Where $c$ is the central charge of the CFT and $a$ is the UV cutoff.
\item For the case when $A$ is a single interval of length $\ell$, and periodic boundary conditions are imposed on the whole system, that is to say, CFT lives on a cylinder. One can find 
\begin{equation}
	S_{A}\sim(c / 3) \log ((L / \pi a) \sin (\pi \ell / L)),
\end{equation}
where $L$ is the circumference of the circle at the base of the cylinder. 
\item { Consider a spatial interval $A$ with length $\ell$ on a plane at a finite temperature $\beta^{-1}.$ The CFT now resides on a cylinder with a compact thermal cycle, i.e., the topology is $M=S^1\times \mathbb{R}.$ The entanglement entropy of $A$ is}
\begin{equation}
	S_{A}\sim (c / 3) \log (( \beta / \pi a) \sinh (\pi \ell / \beta)).
\end{equation}

 \end{itemize}
 
 For higher-dimensional CFT, the calculation of entanglement entropy becomes more complex and challenging. Fortunately, Ryu and Takayanagi~\cite{Ryu:2006bv,Ryu:2006ef,Hubeny:2007xt} proposed that the calculation of entanglement entropy can be transformed into the calculation of the area of a minimal surface in gravity through the AdS/CFT correspondence~\cite{Maldacena:1997re,Witten:1998qj}. Specifically, entanglement entropy $S_A$ in CFT$_{d+1}$ can be computed from the following area law relation
 \begin{equation}
 	S_{A}=\frac{\operatorname{Area}\left(\gamma_{A}\right)}{4 G_{N}^{(d+2)}},\label{6}
 \end{equation}
 where $\gamma_A$ is the minimal surface in $AdS_{d+2}$ with boundary $\partial A$. The area of $\gamma_A$ is denoted as Area$(\gamma_A)$, and $G_N^{(d+2)}$ is the Newton constant in $d+2$-dimensional gravity. When this method is applied to AdS$_3$, Ryu and Takayanagi perfectly reproduced the entanglement entropy in two-dimensional CFT~\cite{Ryu:2006bv}. 
 
 However, proving the Ryu-Takayanagi formula~\eqref{6} or holographic entanglement entropy formula is not an easy task. Fursaev was the first to make some attempts~\cite{Fursaev:2006ih}, but unfortunately there are some problems in his calculations. Subsequently, the authors of~\cite{Casini:2011kv} found that considering the spherical entanglement surface in CFT can map this spherical region to a hyperbolic geometry through conformal transformation. The vacuum state is mapped to the thermal state in the hyperbolic geometry. Then, according to the AdS/CFT correspondence, the entanglement entropy of the original CFT can be calculated by calculating the thermal entropy in gravity. With this idea, the holographic entanglement entropy formula for the special case of a spherical entanglement surface was proven. A more general proof was provided by Lewkowycz and Maldacena~\cite{Lewkowycz:2013nqa}, who employed the Euclidean gravitational path integral method. This approach has become a powerful tool for understanding holographic entanglement entropy and was later extended to demonstrate covariant holographic entanglement entropy~\cite{Dong:2016hjy}. For further information, please see references~\cite{Nishioka:2018khk,Rangamani:2016dms}.
 
It should be noted that Lewkowycz and Maldacena's method appears to rely heavily on the AdS/CFT correspondence and may not be easily extended to other holographic theories. On the other hand, the conformal transformation method in reference~\cite{Casini:2011kv} is more readily applicable to other holographic theories and has been used for entanglement entropy in Warped Conformal Field Theory~\cite{Castro:2015csg}. However, finding the transformation from the vacuum state to the thermal state can be challenging. The problem of finding the transformation from the vacuum state to the thermal state has been solved in the cases where the boundary is a two-dimensional field theory~\cite{Song:2016gtd,Jiang:2017ecm}. The authors of references~\cite{Song:2016gtd,Jiang:2017ecm} named their method of finding this transformation as the generalized Rindler method. This is also the method that will be used in this paper, and we will review it in Section~\ref{3.1}.
 
\begin{figure}[htbp]
	\centering
	\includegraphics[width=0.7\linewidth]{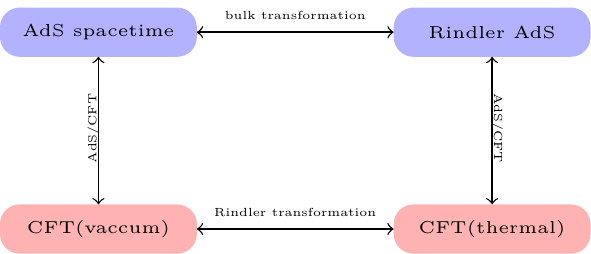}
	\caption{A schematic diagram illustrating the relationship between gravity, field theory, and their connections in the study of holographic entanglement entropy using the generalized Rindler method}
	\label{fig:f1}
\end{figure}

 For the usual AdS$_3$/CFT$_2$, the operation of calculating (or deriving) holographic entanglement entropy using the generalized Rindler method is shown in Fig.~\ref{fig:f1}. 
The process of the operation is as follows:
\begin{itemize}
	\item Calculate the Killing vector field of AdS$_3$. On the asymptotic boundary, the global conformal transformation generators in the boundary CFT can be obtained from the Killing vector fields.
	\item Once the global conformal transformation generators are obtained, the Rindler transformation in the field theory can be calculated using the generalized Rindler method (see Section~\ref{3.1}). This transformation maps the entanglement entropy in the CFT to the thermal entropy in another CFT. By directly calculating the thermal entropy, the entanglement entropy can be obtained. This is also a method for calculating entanglement entropy directly in field theory without relying on holography.
	\item  The bulk transformation can be obtained from the Rindler transformation in field theory according to the generalized Rindler method. The bulk transformation transforms AdS$_3$ into Rindler AdS$_3$. By calculating the black hole entropy of Rindler AdS$_3$ (equivalent to calculating the black hole area), the thermal entropy of the thermal CFT can be obtained. The black hole area is directly related to the area of the minimal surface through the bulk transformation.
\end{itemize}

In this paper, we aim to calculate the entanglement entropy in a CFT where the left-moving central charge is not equal to the right-moving central charge ($c_L \neq c_R$) using the generalized Rindler method. Specifically, we consider a bulk action containing a Chern-Simons term, i.e., topologically massive gravity (TMG), with a CFT on the boundary where $c_L \neq c_R$. The holographic entanglement entropy in this case has been discussed in~\cite{Jiang:2019qvd,Castro:2014tta,Sun:2008uf}, and the Rindler method has also been discussed by the author in~\cite{Jiang:2019qvd}. Here, we will give a more detailed discussion.~\cite{Kraus:2005zm} shows that even when a Chern-Simons term is added to the bulk action, global AdS$_3$ remains a solution to the bulk. Therefore, all our derivations will start from global AdS$_3$. An important feature of the entanglement entropy in a CFT with gravitational anomaly is the presence of a Lorentzian anomaly. In Section~\ref{3.3}, we will show how to derive the anomalous term in the entanglement entropy from the calculation of the thermal entropy in the field theory. For more information on holographic gravitational anomaly and entanglement entropy in a CFT with gravitational anomaly, please refer to~\cite{Castro:2014tta,Kraus:2005zm}.

The organization of this paper is as follows: In Sec. \ref{s2}, we will briefly review the AdS$_3$ spacetime and find out its Killing vector fields, which correspond to the generators of the conformal group of the two-dimensional CFT on the boundary. In Sec. \ref{s3}, we will introduce how to use the Rindler method to calculate the entanglement entropy of a two-dimensional CFT with gravitational anomaly in field theory. In particular, we will carefully deal with {\blue the UV} cutoffs in order to analyze the anomalous contribution of the entanglement entropy in detail. In Sec. \ref{s4}, we will derive the Rindler transformation in the bulk and calculate the thermal entropy of Rindler AdS. In Sec. \ref{s5}, we will derive the entanglement entropy of a zero-temperature CFT on a cylinder and a finite-temperature CFT on a plane. We will give a summary of this paper in Sec. \ref{s6}.

	\section{AdS$_3$ and Killing vector fields}\label{s2}
A fundamental property of AdS spacetime is that the isometry group of a $(d+1)$-dimensional AdS spacetime is identical to the conformal group of a $d$-dimensional Minkowski spacetime. This is a significant finding in the study of AdS/CFT correspondence, which suggests that Killing vector fields in the bulk are related to the generators of global conformal symmetry on the boundary. We start with the flat four-dimensional space $\mathbb{R}^{2,2}$, whose metric is
\begin{equation}
	ds^2=-dy_{0}^{2}-dy_{1}^{2}+dy_{2}^{2}+dy_{3}^{2},\qquad G_{ab}=\text{diag}(-1,-1,1,1).
\end{equation}
One can immediately obtain the Killing vector fields of this metric, which correspond to two types of transformations, namely, rotations and boosts. Their explicit forms are given as follows~\cite{PhysRevD.59.104001,Natsuume:2014sfa,hartman2015lectures}
\begin{eqnarray}
	\text{Rotation:}\qquad L_{a b}&=&y_{a} \partial_{b}-y_{b} \partial_{a},\qquad ab=01,23\\
	\text{Boost:}\qquad J_{a b}&=&y_{a} \partial_{b}+y_{b} \partial_{a},\qquad ab=02,03,12,13
\end{eqnarray}
and these generators form the group $SO(2,2)$. The AdS$_3$ can be embedded into this flat spacetime via 
\begin{equation}
	-L^2=-y^2_0-y^2_1+y^2_2+y^2_3,\label{surface}
\end{equation}
where $L$ is the AdS radius, and we will set it to $L=1$ in the remainder of this paper. As AdS$_3$ possesses $SO(2,2)$ symmetry, we can obtain its Killing vector fields by ``restricting" the previously obtained Killing vector fields to AdS$_3$. The procedure is as follows:  Assume that the manifolds $M=(\text{AdS}_3,g_{ab})$ and $N=(\mathbb{R}^{2,2},G_{ab})$, and suppose there exists a smooth mapping $\phi:M\rightarrow N$ such that for all $ p\in M$, we have $\phi(p)\in N$. If there exists a coordinate system $\{x^\alpha\}\equiv\{x^0,x^1,x^2\}$  in a neighborhood of point $p$ in $M$, and a coordinate system $\{y^\mu\}\equiv\{y^0,y^1,y^2,y^3\}$ in a neighborhood of point $\phi(p)$ in $N$, then the Killing vector fields in $M$ and $N$ are related as follows
\begin{equation}
	\tilde{K}^a=\tilde{K}^{\alpha}\left( \frac{\partial}{\partial x^{\alpha}} \right) ^a=g^{\alpha \beta}\left( \phi ^{\ast}K \right) _{\beta}\left( \frac{\partial}{\partial x^{\alpha}} \right) ^a=g^{\alpha \beta}\frac{\partial y^{\mu}}{\partial x^{\beta}}G_{\mu \nu}K^{\nu}\left( \frac{\partial}{\partial x^{\alpha}} \right) ^a.\label{killing}
\end{equation}
Here, $\tilde{K}^a$ is a Killing vector field ``restricted" in AdS$_3$ and $K^a$ is a Killing vector field in $\mathbb{R}^{2,2}$. The expression $(\phi^\ast K)_a$ represents the pullback of the dual vector $K_a$ under the map $\phi$. We define 
\begin{gather}
		L_{1}=\left(-L_{01}+L_{23}\right) / 2 , \quad L_{2}=\left(J_{12}-J_{03}\right) / 2 , \qquad L_{3}=\left(J_{02}+J_{13}\right) / 2,\\
		\bar{L}_{1}=\left(-L_{01}-L_{23}\right) / 2 , \quad \bar{L}_{2}=\left(J_{12}+J_{03}\right) / 2, \qquad\bar{L}_{3}=\left(J_{02}-J_{13}\right) / 2.
\end{gather}
	One can easily check that:
	\begin{gather}
		\left[L_{1}, L_{2}\right]=-L_{3} ; \quad\left[L_{1}, L_{3}\right]=L_{2} ; \quad\left[L_{2}, L_{3}\right]=L_{1},\\
		\left[ \bar{L}_1,\bar{L}_2 \right] =-\bar{L}_3;\quad \left[ \bar{L}_1,\bar{L}_3 \right] =\bar{L}_2;\quad \left[ \bar{L}_2,\bar{L}_3 \right] =\bar{L}_1.
	\end{gather}
	Then, we redefine
	\begin{gather}
		L_0=L_2,\qquad L_-=L_1+L_3,\qquad L_+=L_1-L_3,\label{10}
		\\
		\bar{L}_0=\bar{L}_2,\qquad \bar{L}_-=\bar{L}_1+\bar{L}_3,\qquad \bar{L}_+=\bar{L}_1-\bar{L}_3.\label{11}
	\end{gather}
One can readily verify that the new Killing vector fields satisfy the Lie algebra of $SL(2,L)$ and $SL(2,R)$, respectively, 
\begin{eqnarray}
	SL(2,L):&\qquad \left[ L_0,L_{\pm} \right] =\mp L_{\pm},\qquad \left[ L_+,L_- \right] =2L_0,\\ 
	SL(2,R):&\qquad \left[ \bar{L}_0,\bar{L}_{\pm} \right] =\mp \bar{L}_{\pm},\qquad \left[ \bar{L}_+,\bar{L}_- \right] =2\bar{L}_0.
\end{eqnarray}
	By making the following coordinate transformation, we can relate the hypersurface defined by \eqref{surface} to the Poincar\'e coordinates in the AdS$_{3}$ spacetime~\cite{Casini:2011kv}:
	\begin{equation}
		y_0+y_3=\frac{L^2}{z},\qquad  y_1=\frac{L}{z}t,\qquad y_2=\frac{L}{z}x.\label{14}
	\end{equation}
	Together with the defining equation of the hypersurface \eqref{surface}, we totally have four equations, from which we can solve for $y^\mu\equiv y_\mu$  in terms of $y_\mu(x^\alpha)$. To facilitate later calculations, we perform another coordinate transformation in this paper:
	\begin{equation}
		u=x+t,\qquad v=x-t.\label{15}
	\end{equation}
The induced metric on the hypersurface is:
\begin{equation}
	ds^2=\frac{1}{z^2}\left( dz^2+dudv \right) .\label{16}
\end{equation}
	Taking the asymptotic limit as $z\rightarrow 0$ and removing the $1/z^2$ factor, we obtain the metric for the dual CFT living on the boundary. Using equation \eqref{killing}, we can obtain the Killing vector fields in AdS$_{3}$ that corresponds to the Killing vector fields given in equations \eqref{10} and \eqref{11}, namely:
	\begin{eqnarray}
		\tilde{L}_0&=&u\partial _u+\frac{z}{2}\partial _z,\qquad\tilde{L}_+=-\partial _u,\qquad\tilde{L}_-=-u^2\partial _u+z^2\partial _v-uz\partial _z,
		\\
		\tilde{\bar{L}}_0&=&-v\partial _v-\frac{z}{2}\partial _z,\qquad\tilde{\bar{L}}_+=-z^2\partial _u+v^2\partial _v+vz\partial _z,\qquad\tilde{\bar{L}}_-=\partial _v.
	\end{eqnarray}
	On the asymptotic boundary, the global generators of the conformal transformation are
	\begin{gather}
		l_0=u\partial _u,\qquad l_+=-\partial _u,\qquad l_-=-u^2\partial _u,\label{19}
		\\
		\bar{l}_0=-v\partial _v,\qquad \bar{l}_+=v^2\partial _v,\qquad\bar{l}_-=\partial _v.\label{20}
	\end{gather}
	With these ingredients, we can use the generalized Rindler method~\cite{Casini:2011kv} to solve the holographic entanglement entropy in AdS$_3$/CFT$_2$.

\section{Entanglement entropy from field theory}\label{s3}
     In this section, we will investigate the entanglement entropy of a two-dimensional CFT with gravitational anomalies using the generalized Rindler method. To this end, we will first provide a brief overview of the Rindler method. For detailed information, please refer to references~\cite{Casini:2011kv,Castro:2015csg,Song:2016gtd,Jiang:2017ecm}.
\subsection{(Generalized) Rindler method}\label{3.1}

    The main idea of the Rindler method is to use a symmetry transformation to transform the calculation of entanglement entropy in field theory to the calculation of thermal entropy. In the context of holography, the thermal entropy of the boundary field theory is related to the black hole entropy in the bulk. Therefore, the entanglement entropy in field theory can be calculated by computing the black hole entropy in gravity from the Rindler method. The key to the Rindler method is to find the appropriate Rindler transformation. The main approach to find the Rindler transformation has been summarized in reference~\cite{Jiang:2017ecm}, and we will provide a brief introduction here.

    Assuming that the quantum field theory on the manifold $B$ is invariant under the group $G$, and the generators of the global symmetries in the group $G$ are denoted by $h_i$. If we want to calculate the entanglement entropy of a subregion $A$, then the Rindler transformation $R$ is a symmetry transformation that maps the causal domain of $A$, which is denoted as $\mathcal{D}$, to a non-compact manifold $\tilde{B}$. The Rindler transformation $R$ here should satisfy:
\begin{enumerate}
	\item 
	 Rindler transformation 
		\begin{equation}
		\left. \begin{array}{c}
			R:\mathcal{D} \rightarrow \tilde{B}\\
			\quad~ x\mapsto \tilde{x}\\
		\end{array} \right. 
	\end{equation}
	 is a symmetry transformation. Here $x=\left( x^0,x^1 \right) \in \mathcal{D} $, $\tilde{x}=\left( \tilde{x}^0,\tilde{x}^1 \right) \in \tilde{B}$.
	\item The new coordinates should satisfy $\tilde{x}^i\sim\tilde{x}^i+i\tilde\beta^i$ in order to obtain a thermal state. 
	\item The vectors $\partial_{\tilde{x}^i}$ annihilate vacuum, that is
	\begin{equation}
		\partial_{\tilde{x}^{i}}=\sum_{j} b_{i j} h_{j},\label{21}
	\end{equation}
where $b_{ij}$ are arbitrary constants.
\item Let $k_t = \tilde \beta^i\partial_{\tilde x^{i}}$ be the generator of the modular flow, such that the causal domain remains invariant under this flow for the selected region. 
\end{enumerate}
	Under these conditions, the Rindler transformation can be derived. In the next subsection, we will apply the Rindler method to provide a detailed calculation of the entanglement entropy in field theory. The holographic dictionary tells us that the global generators of the asymptotic symmetry group correspond to the isometry generators of the bulk metric. By substituting the generators of the bulk isometries for the $h_i$ generators, and requiring the Rindler bulk space to satisfy the same boundary conditions, we can obtain the Rindler transformations in the bulk spacetime.
	\subsection{Rindler transformation in CFT$_2$}
	In this section, we will demonstrate how to use the Rindler method to calculate the entanglement entropy of a subsystem in a CFT. This calculation can also be found in references\cite{Song:2016gtd,Jiang:2017ecm,Wen:2018whg}, but our presentation will be more detailed than previous works.
	
	Consider a region $\mathcal{I}$ in a zero-temperature CFT on a plane, which can be described by the following equation:
	\begin{equation}
		\mathcal{I} \equiv \left\{ \left( u,v \right) \bigg|u=l_u\left( s-\frac{1}{2} \right) ,v=l_v\left( s-\frac{1}{2} \right) ,s\in \left[ 0,1 \right] \right\} ,\label{22}
	\end{equation}
where $l_u$ and $l_v$ are some positive constants which restrict the domain of dependence (DOD) of $\mathcal{I}$ to 
	\begin{equation}
		\mathcal{D} =\left\{ \left( u,v \right) \bigg|-\frac{l_u}{2}<u<\frac{l_u}{2},-\frac{l_v}{2}<v<\frac{l_v}{2} \right\}.
\end{equation}
According to the Rindler method, we can map this region to a thermal state via a conformal transformation. Assuming that this conformal transformation is
	\begin{equation}
		u^\prime=f(u),\qquad v^\prime=g(v).
	\end{equation}
	In the field theory, the Rindler transformation is 
	\begin{gather}
		u^\prime=\frac{\beta_{u^\prime}}{\pi}\arctan\!\text{h}\frac{2u}{l_u},\label{31}\\
			v^\prime=-\frac{\beta_{v^\prime}}{\pi}\arctan\!\text{h}\frac{2v}{l_v}.\label{32}
	\end{gather}
	 The specific derivation process is given in Appendix \ref{A}.
 If we take  $\beta_{u^\prime}=-\beta_{v^\prime}=\pi$, then our Rindler transformation is consistent with the results of reference~\cite{Wen:2018whg}. On the other hand, if we set $l_u=l_v=2R$, then we have
	\begin{gather}
		u=R\tanh u^{\prime}=R\frac{1-e^{-2u^{\prime}}}{1+e^{-2u^{\prime}}},\\
		v=R\tanh v^{\prime}=R\frac{1-e^{-2v^{\prime}}}{1+e^{-2v^{\prime}}}.
	\end{gather}
   This is formally the same as the equation $(2.31)$ in reference~\cite{Casini:2011kv}, which represents the coordinate transformation that maps the vacuum state to the thermal state. The reason for the lack of exact equivalence is because we have taken $\beta_{u^\prime}=-\beta_{v^\prime}=\pi$ instead of $\beta_{u^\prime}=-\beta_{v^\prime}=2\pi$ in the reference~\cite{Casini:2011kv}.
   
\begin{figure}[htb]
   	\centering
   	\includegraphics[width=0.8\linewidth]{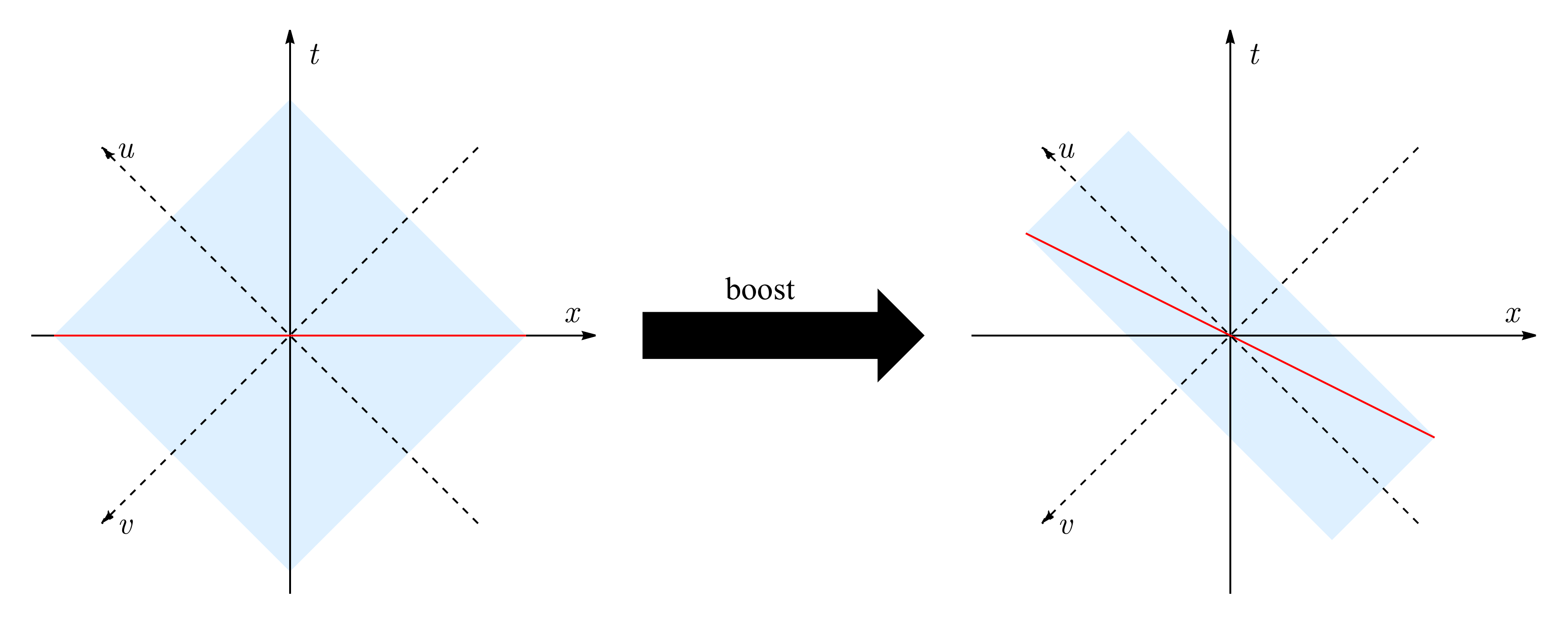}
   	\caption{A schematic diagram of an interval (in red) undergoing a boost.}
   	\label{fig:boost}
   \end{figure}

   The computation of entanglement entropy in CFT yields a divergent result, and we need to introduce an appropriate cutoff, i.e., a regularization scheme, to compute the entanglement entropy of the region under consideration. Assume the region obtained by truncating $\mathcal{I}$ as
  \begin{equation}
  	   \mathcal{I} _{\mathrm{reg}}\equiv \left\{ \left. \left( u,v \right) \right.\bigg|u=l_u\left( s-\frac{1}{2} \right) ,v=l_v\left( s-\frac{1}{2} \right) ,s\in \left[ \varepsilon _1,1-\varepsilon _2 \right] \right\} .
  \end{equation}
  After the Rindler transformation similar to the above procedures, $\mathcal{I} _{\mathrm{reg}}$ becomes $\mathcal{I}^\prime_{\text{reg}}$: 
  \begin{equation}
  	\mathcal{I} _{\mathrm{reg}}^{\prime}= \left\{ \left. \left( u^{\prime},v^{\prime} \right) \right.\bigg|u^{\prime}=\frac{\beta _{u^{\prime}}}{\pi}\arctan\!\text{h} \left[ 2\left( s-\frac{1}{2} \right) \right] ,v^{\prime}=-\frac{\beta _{v^{\prime}}}{\pi}\arctan\!\text{h} \left[ 2\left( s-\frac{1}{2} \right) \right] ,s\in \left[ \varepsilon _1,1-\varepsilon _2 \right] \right\} ,\label{57}
  \end{equation}
   where $\varepsilon_1,\varepsilon_2\ll 1$ are the cutoffs. For the convenience, we set
   \begin{gather}
   	\varepsilon _{u1}=l_u\varepsilon _1,\qquad \varepsilon _{u2}=l_u\varepsilon _2,\label{55}
   	\\
   	\varepsilon _{v1}=l_v\varepsilon _1,\qquad \varepsilon _{v2}=l_v\varepsilon _2.\label{56}
   \end{gather}
 Then we have
\begin{equation}
\frac{\varepsilon _{u1}}{\varepsilon _{v1}}=\frac{\varepsilon _{u2}}{\varepsilon _{v2}}=\frac{l_u}{l_v}.
\end{equation}
   This equation reflects the degree to which the interval we consider deviates from the equal time surface (when the interval is on the equal time surface, $l_u/l_v=1$). Perhaps a more accurate phrasing would be that $l_u/l_v$ measures the boost of the DOD. We can convert it to the form of boost angle. To this end, we consider a boost that induces a Lorentz transformation (refer to Fig. \ref{fig:boost}),
   \begin{equation}
   	\begin{aligned}
   		t^{\prime}&=\gamma \left( t-\nu x \right) =\frac{1}{\sqrt{1-\nu ^2}}\left( t-\nu x \right) ,\\
   		x^{\prime}&=\gamma \left( x-\nu t \right) =\frac{1}{\sqrt{1-\nu^2}}\left( x-\nu t \right) .\\
   	\end{aligned}
   \end{equation}
 where   $\nu$ is the velocity of the boosted interval relative to the selected inertial frame. Under such a configuration, it can be calculated that 
   \begin{equation}
   	\frac{l_u}{l_v}=e^{-2\kappa}
   \end{equation}
   for the boosted interval, where $\kappa=\arctan\!\text{h}\nu$ is the boost angle.

    Here is something that needs to be stressed. The Rindler transformations \eqref{31} and \eqref{32} are derived for a specific interval $\mathcal{I}$, but they are also applicable to any spacelike curves that have the same DOD. For such a general curve, equations \eqref{55} and \eqref{56} are not satisfied.
   \begin{figure}[htb]
   	\centering
   	\includegraphics[width=0.5\linewidth]{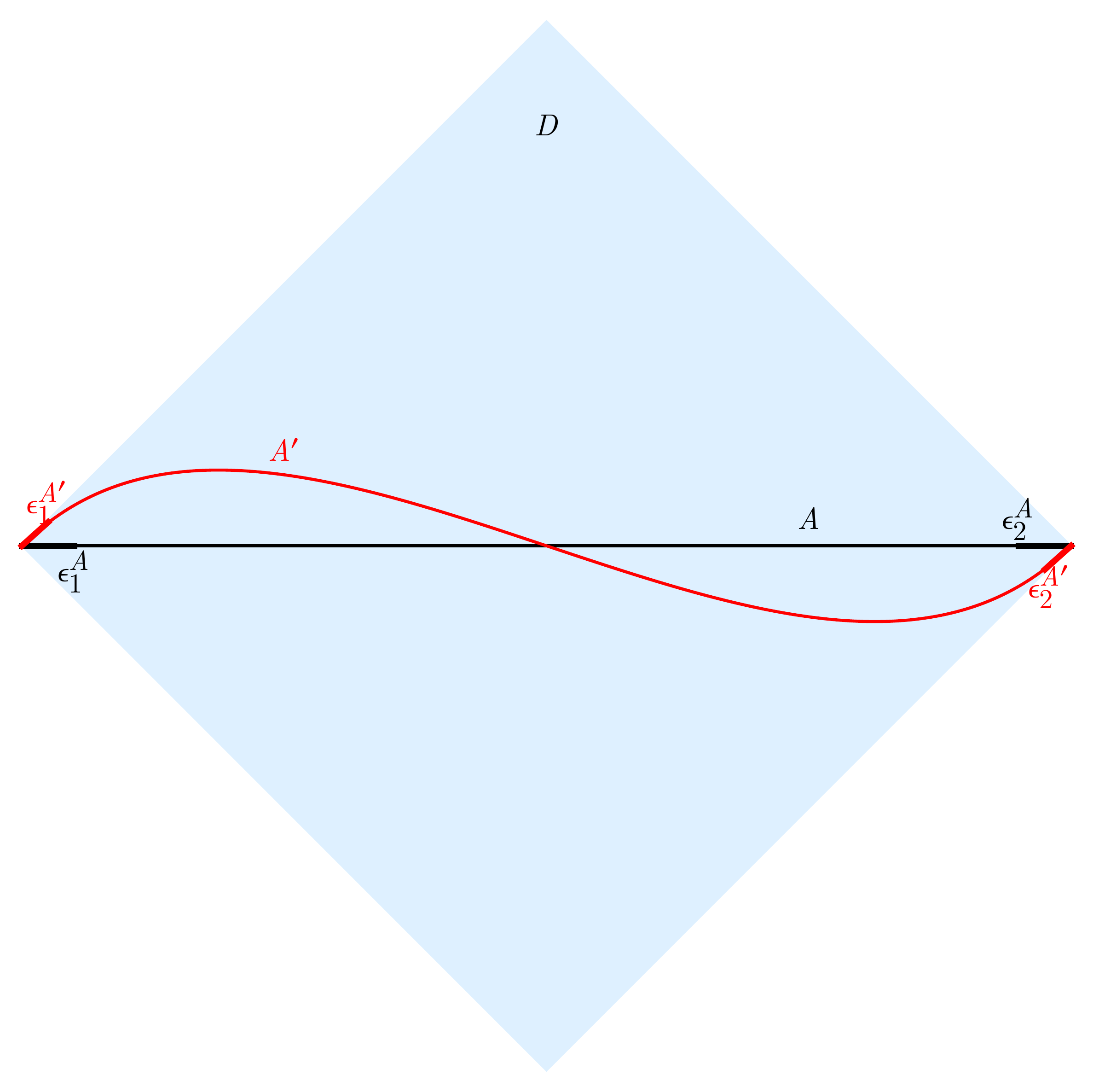}
   	\caption{A schematic diagram of different spacelike curves $A$ and $A'$ having the same DOD. The $\epsilon^{A,A'}_{\#}$'s represent the small truncations at the endpoints of $A$ and $A'$ respectively. }
   	\label{fig:two-interval}
   	\end{figure}
   	As shown in the Fig. \ref{fig:two-interval}, there are two spacelike curves, $A$ and $A^\prime$, having a common  DOD $D$. The $ \epsilon^{A,A'}_{\#}$'s in the figure are the truncations at the endpoints of $A$ and $A'$. For convenience, let's assume that $\epsilon^A_1$ and $\epsilon^A_2$ are boosted to coincide with $\epsilon^{A^\prime}_1$ and $\epsilon^{A^\prime}_2$ respectively. We can decompose the truncations of $A^\prime$ into truncations in the $u$ and $v$ directions and denoted them as $\epsilon^{A^\prime}_{u1},\epsilon^{A^\prime}_{v1},\epsilon^{A^\prime}_{u2},\epsilon^{A^\prime}_{v2}$. They satisfy the following relationship:
   	\begin{equation}
   	   	\begin{aligned}
   		\frac{\epsilon _{u1}^{A^{\prime}}}{\epsilon _{v1}^{A^{\prime}}}&=e^{-2\kappa _1},\qquad \frac{\epsilon _{u2}^{A^{\prime}}}{\epsilon _{v2}^{A^{\prime}}}=e^{-2\kappa _2},\\
   		\epsilon _{u1}^{A^{\prime}}\epsilon _{v1}^{A^{\prime}}&=\left( \epsilon _{1}^{A} \right) ^2,\qquad \epsilon _{u2}^{A^{\prime}}\epsilon _{v2}^{A^{\prime}}=\left( \epsilon _{2}^{A} \right) ^2,
   	\end{aligned}
   	\end{equation}
   	where $\kappa_1$ and $\kappa_2$ are boost angles of $\epsilon^{A^\prime}_{1}$ and $\epsilon_{2}^{A^\prime}$, respectively.

	\subsection{Entanglement entropy of zero temperature CFT on the plane}\label{3.3}
\begin{figure}[htb]
	\centering
	\includegraphics[width=0.8\linewidth]{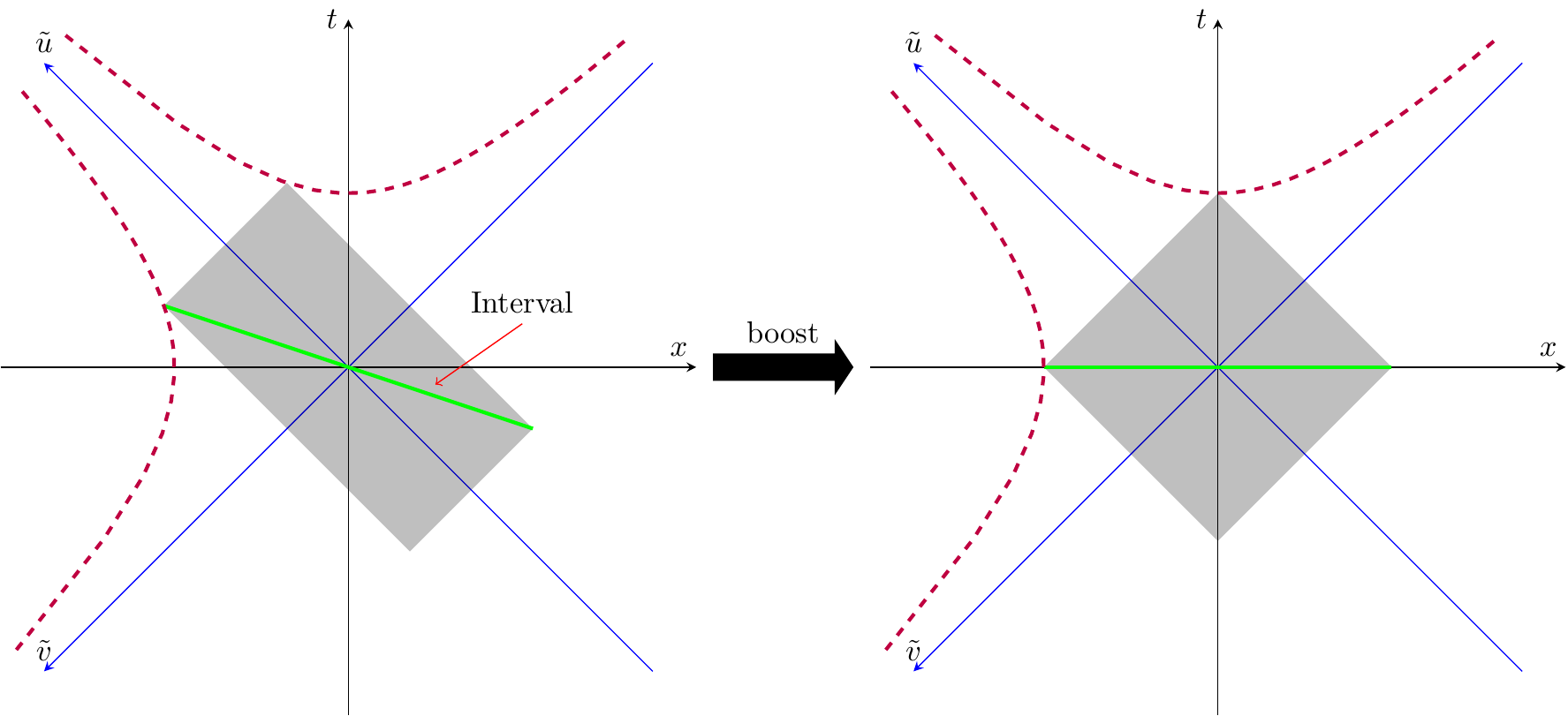}
	\caption{A diagram illustrating the boost transformation of an interval's DOD into a DOD on an equal-time surface at $t=0$. }
	\label{fig:domain}
\end{figure}

	Consider a CFT living on an arbitrary torus with the following identification
	\begin{equation}
\left( \tilde{u},\tilde{v} \right) \sim \left( \tilde{u}+i\bar{a},\tilde{v}-ia \right) \sim \left( \tilde{u}+2\pi \bar{b},\tilde{v}-2\pi b \right)  ,
	\end{equation}
	where  $\bar{a},a,\bar{b},b$ are some real constants. Without loss of generality, we assume $\bar{a}, \bar{b}>0$. The partition function of the CFT on this torus can be written as $ Z_{\bar{b}|b}\left( \bar{a}|a \right)  $. It is worth noting that we can always use a Lorentz transformation to turn such an identification into a more ``standard" identification. The reason can be briefly summarized as follows: an arbitrary causal region, as shown in the left panel of Fig.~\ref{fig:domain}, can always be transformed into a diamond-shaped region on an equal-time surface through a boost. We may take this equal-time surface to be the $t=0$ surface.
Assuming that the coordinates of the leftmost point of the causal region in the left panel of Fig.\ref{fig:domain} are $(\pi \bar{b},\pi |b|)$, then under the boost this point should move along the curve 
\begin{equation}
\tilde{u}=\frac{\pi ^2|b|\bar{b}}{\tilde{v}}.
\end{equation}
The coordinates of the left endpoint in the right panel of Fig.\ref{fig:domain}  should be $(\pi\sqrt{|b|\bar b},\pi\sqrt{|b|\bar b})$. Setting 
\begin{equation}
	\tilde{u}^{\prime}=\sqrt{\frac{|b|}{\bar{b}}}\tilde{u},\qquad \tilde{v}^{\prime}=\sqrt{\frac{\bar{b}}{|b|}}\tilde{v}\label{62}
\end{equation}
 is equivalent to performing a boost that transforms a boosted DOD into an unboosted DOD. Thus, we can consider the CFT on the following torus
\begin{equation}
\left( \tilde{u}^{\prime},\tilde{v}^{\prime} \right) \sim \left( \tilde{u}^{\prime}+i\sqrt{\frac{\left| b \right|}{\bar{b}}}\bar{a},\tilde{v}^{\prime}-i\sqrt{\frac{\bar{b}}{\left| b \right|}}a \right) \sim \left( \tilde{u}^{\prime}+2\pi \sqrt{\left| b \right|\bar{b}},\tilde{v}^{\prime}-2\pi \operatorname{sgn} \left( b \right) \sqrt{\left| b \right|\bar{b}} \right)  ,
\end{equation}
where ``$\operatorname{sgn}$" is the sign function. Then we have
\begin{equation}
	Z_{\bar{b}|b}\left( \bar{a}|a \right) =Z_{\sqrt{\left| b \right|\bar{b}}|\operatorname{sgn}(b)\sqrt{\left| b \right|\bar{b}}}\left( \sqrt{\frac{\left| b \right|}{\bar{b}}}\bar{a}|\sqrt{\frac{\bar{b}}{\left| b \right|}}a \right) .
\end{equation}
If we make a scaling transformation
\begin{equation}
	\hat{u}=\frac{\tilde{u}^{\prime}}{\sqrt{\left| b \right|\bar{b}}},\qquad \hat{v}=\frac{\tilde{v}^{\prime}}{\sqrt{\left| b \right|\bar{b}}},
\end{equation}
we can obtain the usual canonical circle
\begin{equation}
\left( \hat{u},\hat{v} \right) \sim \left( \hat{u}+i\frac{\bar{a}}{\bar{b}},\hat{v}-i\frac{a}{\left| b \right|} \right) \sim \left( \hat{u}+2\pi ,\hat{v}-2\pi \text{sgn} \left( b \right) \right)  ,\label{66}
\end{equation}
 where the spatial period is $2\pi$. This is still a conformal transformation and the partition function will not change. 
 On the equal-time surface at $t=0$, we can calculate the partition function as in the usual CFT by considering $\hat u=\bar{z},\hat v=z$, so
\begin{equation}
\begin{aligned}
Z_{\bar{b}|b}\left( \bar{a}|a \right) &=Z_{1|\text{sgn} \left( b \right)}\left( \frac{\bar{a}}{\bar{b}}|\frac{a}{\left| b \right|} \right)\\
 &=\mathrm{Tr}_{1|\text{sgn} \left( b \right)}\left( e^{-\frac{\bar{a}}{\bar{b}}\bar{\mathcal{L}}_{0}^{\mathrm{Cyl}}}e^{\frac{a}{\left| b \right|}\mathcal{L} _{0}^{\mathrm{Cyl}}} \right) =\mathrm{Tr}_{1|\text{sgn} \left( b \right)}\left( e^{-\frac{\bar{a}}{\bar{b}}\left( \bar{\mathcal{L}}_0-\frac{c_L}{24} \right)}e^{\frac{a}{\left| b \right|}\left( \mathcal{L} _0-\frac{c_R}{24} \right)} \right) ,
\end{aligned}
\end{equation}
where $\bar{\mathcal{L}}^{\text{Cyl}}_0,\mathcal{L}_0^\text{Cyl}$ are  the charge of $\partial_{\hat u}$ and $\partial_{\hat v}$ on the cylinder with the canonical spatial cycle while $\bar{\mathcal{L}}_0,\mathcal{L}_0$ are the charge on a plane. The left-moving central charge $c_L$ and right-moving central charge $c_R$ of a CFT are generally not equal.  The relationship between the second and third equal signs can be found in~\cite{Blumenhagen:2009zz}. Next, we make a conformal transformation to exchange the spatial circle and thermal circle, which is called the S-transformation~\cite{hartman2015lectures,DiFrancesco:1997nk,Blumenhagen:2009zz}
\begin{equation}
	\hat{u}^{\prime}=2\pi i\frac{\hat{u}}{\beta _{\hat{u}}},\qquad \hat{v}^{\prime}=2\pi i\text{sgn} \left( b \right) \frac{\hat{v}}{\beta _{\hat{v}}},
\end{equation}
 where we have let $\beta _{\hat{u}}=\frac{\bar{a}}{\bar{b}},\beta _{\hat{v}}=\frac{a}{|b|}$. From this we obtain
 \begin{equation}
\left( \hat{u},\hat{v} \right)\sim \left( \hat{u}-2\pi ,\hat{v}+2\pi \text{sgn} \left( b \right) \right) \sim \left( \hat{u}+i\frac{4\pi^2}{\beta_{\hat{u}}},\hat{v}-i\frac{4\pi^2}{\beta_{\hat{v}}} \right).
 \end{equation}
This is equivalent to replacing $\beta_{\hat{u}}$ with $4\pi^2/\beta_{\hat{u}}$ and $\beta_{\hat{v}}$ with $4\pi^2/\beta_{\hat{v}}$ in~\eqref{66}. Then we have
\begin{equation}
Z_{1|\text{sgn} \left( b \right)}\left( \frac{\bar{a}}{\bar{b}}|\frac{a}{\left| b \right|} \right) =Z_{1|\text{sgn} \left( b \right)}\left( \beta _{\hat{u}}|\beta _{\hat{v}} \right) =Z_{-1|-\text{sgn}(b)}\left( \frac{4\pi ^2}{\beta _{\hat{u}}}|\frac{4\pi ^2}{\beta _{\hat{v}}} \right)  .
\end{equation}
	Assume that the contribution of the vacuum to the partition function is dominant, then we arrive
	{ \begin{equation}
\begin{aligned}
	Z_{1|\text{sgn}(b)}\left( \beta _{\hat{u}}|\beta _{\hat{v}} \right) &=Z_{-1|-\text{sgn}(b)}\left( \frac{4\pi ^2}{\beta _{\hat{u}}}|\frac{4\pi ^2}{\beta _{\hat{v}}} \right) \\
	&=\mathrm{Tr}_{-1|-\text{sgn}(b)}\left( e^{-\frac{4\pi ^2}{\beta _{\hat{u}}}\left( \bar{\mathcal{L}}_0-\frac{c_L}{24} \right)}e^{\frac{4\pi ^2}{\beta _{\hat{v}}}\left( \mathcal{L} _0-\frac{c_R}{24} \right)} \right)
	\approx e^{\frac{\pi ^2c_L}{6\beta _{\hat{u}}}}e^{-\frac{\pi ^2c_R}{6\beta _{\hat{v}}}}.
\end{aligned}
	\end{equation}}
Here we have assumed that the vacuum charge on the plane vanishes. Therefore, we can obtain the thermal entropy
\begin{equation}
\begin{aligned}
S_{\bar{b}|b}\left( \bar{a}|a \right) &=S_{1|\text{sgn}(b)}\left( \beta _{\hat{u}}|\beta _{\hat{v}} \right)
=\left( 1-\beta _{\hat{u}}\partial _{\beta _{\hat{u}}}-\beta _{\hat{v}}\partial _{\beta _{\hat{v}}} \right) \log Z_{1|\text{sgn}(b)}\left( \beta _{\hat{u}}|\beta _{\hat{v}} \right)\\
&=\left( 1-\beta _{\hat{u}}\partial _{\beta _{\hat{u}}}-\beta _{\hat{v}}\partial _{\beta _{\hat{v}}} \right) \left( \frac{\pi ^2c_L}{6\beta _{\hat{u}}}-\frac{\pi ^2c_R}{6\beta _{\hat{v}}} \right)\\
&=\frac{\pi ^2}{3}\frac{c_L}{\beta _{\hat{u}}}-\frac{\pi ^2}{3}\frac{c_R}{\beta _{\hat{v}}}.\label{72}\\
\end{aligned}
\end{equation}
It should be noted that our formula is only valid when
\begin{equation}
	\beta _{\hat{u}}\rightarrow 0^+,\qquad \beta _{\hat{v}}\rightarrow 0^-.
\end{equation}
 The most general formula for the thermal entropy of CFT on an arbitrary torus can be obtained through the same operation,
\begin{equation}
	S_{\bar{b}|b}\left( \bar{a}|a \right) =\frac{\pi ^2}{3}\frac{c_L}{\beta _{\hat{u}}}+\frac{\pi ^2}{3}\frac{c_R}{\beta _{\hat{v}}},\label{3.40}
\end{equation}
where $\beta_{\hat{u}}=\left |\bar{a}/\bar{b}\right |$, $\beta_{\hat{v}}=\left |a/b\right |$. The equation \eqref{3.40}, known as the Cardy formula~\cite{Cardy:1986ie,Hartman:2014oaa}, applies to the case when $\beta_{\hat{u}}\rightarrow 0$ and $\beta_{\hat{v}}\rightarrow 0$.
	
	From~\eqref{57}, we obtain
 \begin{eqnarray}
		\left| \bar{a} \right|&=&\left| \beta _{u^{\prime}} \right|,\qquad\qquad\qquad\qquad \left| a \right|=\left| \beta _{v^{\prime}} \right|,\\
		2\pi \left| \bar{b} \right|&=&\left| \frac{\beta _{u^{\prime}}}{2\pi}\log \frac{\varepsilon _{u1}\varepsilon _{u2}}{l_{u}^{2}} \right|,\qquad 2\pi \left| b \right|=\left| \frac{\beta _{v^{\prime}}}{2\pi}\log \frac{\varepsilon _{v1}\varepsilon _{v2}}{l_{v}^{2}} \right|.\label{59}
\end{eqnarray}
in which we have expanded truncations as small quantities in \eqref{59}. Then, we can obtain
\begin{equation}
\beta _{\hat{u}}=\left| \frac{\bar{a}}{\bar{b}} \right|=-\frac{4\pi ^2}{\log \frac{\varepsilon _{u1}\varepsilon _{u2}}{l_{u}^{2}}},\qquad \beta _{\hat{v}}=\left| \frac{a}{b} \right|=-\frac{4\pi ^2}{\log \frac{\varepsilon _{v1}\varepsilon _{v2}}{l_{v}^{2}}}.
\end{equation}
Substituting it into~\eqref{72} yields the entanglement entropy
\begin{equation}
	S=\frac{c_L}{12}\log \frac{l_{u}^{2}}{\varepsilon _{u1}\varepsilon _{u2}}+\frac{c_R}{12}\log \frac{l_{v}^{2}}{\varepsilon _{v1}\varepsilon _{v2}}.\label{76}
\end{equation}
 It can be rewritten as
\begin{equation}
	S=\frac{c_L+c_R}{24}\log \frac{l_{u}^{2}l_{v}^{2}}{\varepsilon _{u1}\varepsilon _{v1}\varepsilon _{u2}\varepsilon _{v2}}+\frac{c_L-c_R}{24}\log \frac{l_{u}^{2}\varepsilon _{v1}\varepsilon _{v2}}{l_{v}^{2}\varepsilon _{u1}\varepsilon _{u2}}.\label{3.49}
\end{equation}
This result should be applicable to the calculation of the entanglement entropy of all spacelike curves that have the same DOD as $\mathcal I$. It is clear that when $c_L=c_R$, the second term in equation \eqref{3.49} is equal to zero. However, when $c_L\neq c_R$, the second term may contribute. 
\begin{figure}[htb]
	\centering
	\includegraphics[width=0.5\linewidth]{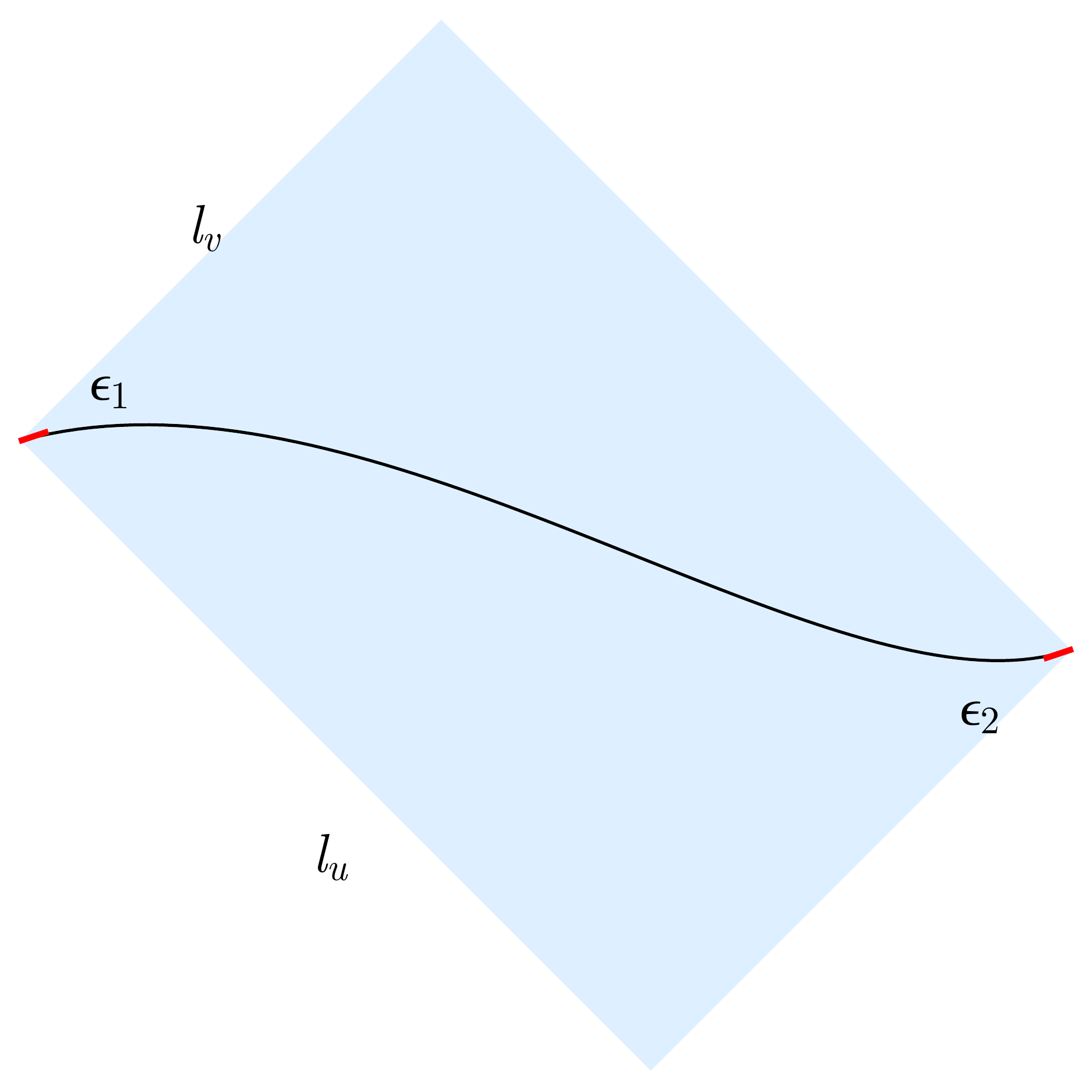}
	\caption{An arbitrary spacelike curve with a boosted DOD. The small red lines are the truncations $\epsilon_1$ and $\epsilon_2$ at the endpoints of the curve. }
	\label{fig:general-case}
\end{figure}
The most general case is that the two truncations have a boost relative to the examined DOD, as shown in Fig. \ref{fig:general-case}. Consider an arbitrary spacelike curve, whose shape and size of the DOD are characterized by $l_u$ and $l_v$, and $\epsilon_1$, $\epsilon_2$ are cutoffs. Assume that
\begin{equation}
	l_u l_v=(2R)^2,\qquad\frac{l_u}{l_v}=e^{-2\kappa},\qquad \epsilon _{1}^{2}=\epsilon _{u1}\epsilon _{u2},\qquad \epsilon _{2}^{2}=\epsilon _{u2}\epsilon _{v2},
\end{equation}
where
\begin{equation}
	\frac{\epsilon _{u1}}{\epsilon _{v1}}=e^{-2\kappa _1},\qquad \frac{\epsilon _{u2}}{\epsilon _{v2}}=e^{-2\kappa _2}.
\end{equation}
According to \eqref{3.49}, we have
\begin{equation}
	S=\frac{c_L+c_R}{12}\log \frac{(2R)^2}{\epsilon _1\epsilon _2}+\frac{c_L-c_R}{12}\left( \kappa _1+\kappa _2-2\kappa \right) .
\end{equation}
furthermore, we can set 
\begin{equation}
	\delta \kappa _1=\kappa _1-\kappa ,\qquad \delta \kappa _2=\kappa _2-\kappa ,
\end{equation}
 which are the boost angles of the cutoffs relative to the DOD. This means that when the total boost angle relative to the DOD vanishes, the anomalous contribution disappears. Otherwise, anomalous contributions will contribute, which have been discussed in some previous literatures \cite{Castro:2014tta,Wall:2011kb,Iqbal:2015vka}.

\section{Entanglement entropy from gravity}\label{s4}
\subsection{Rindler transformation in the bulk and Rindler AdS spacetime}	
In the rest of the article, for convenience, we always take $\beta_{u^\prime}=-\beta_{v^\prime}=\pi$, which will not affect the results. According to the Rindler method, in the bulk we have
\begin{eqnarray}
		\partial _{u^{\prime}}&=&a_0\tilde{L}_0+a_+\tilde{L}_++a_-\tilde{L}_-
		=-\frac{l_u}{2}\tilde{L}_++\frac{2}{l_u}\tilde{L}_- \nonumber\\
		&=&\left( \frac{l_u}{2}-\frac{2}{l_u}u^2 \right) \partial _u+\frac{2}{l_u}z^2\partial _v-\frac{2}{l_u}uz\partial _z,\\
	\partial _{v^{\prime}}&=&b_0\tilde{\bar{L}}_0+b_+\tilde{\bar{L}}_++b_-\tilde{\bar{L}}_-  \nonumber\\
	&=&\frac{2}{l_v}z^2\partial _u+\left( \frac{l_v}{2}-\frac{2}{l_v}v^2 \right) \partial _v-\frac{2}{l_v}vz\partial _z.
\end{eqnarray}
 These two equations were obtained by directly replacing $l_i$ in \eqref{24} and \eqref{33} with $L_i$ and replace $\bar{l}_i$ with $\bar{L}_i$\footnote{Here $l_i$ refers to $l_0, l_+$ and $l_-$ while $L_i$ refers to $L_0, L_+, L_-$. The same meaning goes to $\bar{l}_i$ and $\bar{L}_i$.}.  After the Rindler transformation in the bulk, we can get the metric components from the AdS spacetime~\eqref{16} as
\begin{eqnarray}
		g_{u^{\prime}u^{\prime}}&=&g_{ab}\left( \frac{\partial}{\partial u^{\prime}} \right) ^a\left( \frac{\partial}{\partial u^{\prime}} \right) ^b=1,
		\qquad g_{v^{\prime}v^{\prime}}=g_{ab}\left( \frac{\partial}{\partial v^{\prime}} \right) ^a\left( \frac{\partial}{\partial v^{\prime}} \right) ^b=1,\\
		g_{u^{\prime}v^{\prime}}&=&g_{v^{\prime}u^{\prime}}=g_{ab}\left( \frac{\partial}{\partial u^{\prime}} \right) ^a\left( \frac{\partial}{\partial v^{\prime}} \right) ^b=\frac{ l_{u}^{2}\left( l_{v}^{2}-4v^2 \right) +4\left( -l_{v}^{2}u^2+4\left( uv+z^2 \right) ^2 \right) }{8l_ul_vz^2}.
\end{eqnarray}
Since $\partial_{u^\prime}$ and $\partial_{v^\prime}$ are the Killing vector fields, the metric components should be independent of $u^\prime,v^\prime$. We assume that $g_{u^\prime v^\prime}=F(z^\prime)$ in which $F(z')$ has good enough properties so that we can perform the following operations,
 \begin{equation}
 	z^\prime=F^{-1}(g_{u^\prime v^\prime})\label{67}
 \end{equation}
	and 
	\begin{eqnarray}
		\left( dz^{\prime} \right) ^a&=&\frac{-l_{v}^{2}u+4v\left( uv+z^2 \right)}{l_ul_vz^2F^{\prime}\left( z^{\prime} \right)}\left( du \right) ^a+\frac{-l_{u}^{2}v+4u\left( uv+z^2 \right)}{l_ul_vz^2F^{\prime}\left( z^{\prime} \right)}\left( dv \right) ^a\nonumber\\
		&&+\frac{- \left( l_{u}^{2}-4u^2 \right) \left( l_{v}^{2}-4v^2 \right) +16z^4}{4l_ul_vz^3F^{\prime}\left( z^{\prime} \right)}\left( dz \right) ^a,
	\end{eqnarray}
	where $F^\prime\left (z'\right )$ represents the derivative with respect to $z'$. Then we have
	\begin{equation}
		g^{z^\prime z^\prime}=g^{ab}(dz^\prime)_a (dz^\prime)_b=\frac{4(F(z^\prime)^2-1)}{F^\prime(z^\prime)^2}.
	\end{equation}
	It is worth noting that Rindler AdS is not unique {\blue }and we expect that the metric component $
	g_{u^{\prime}z^{\prime}}=g_{v^{\prime}z^{\prime}}=0
	$.  Therefore, we reach
	\begin{gather}
		g_{z^\prime z^\prime}=\frac{F^\prime(z^\prime)^2}{4(F(z^\prime)^2-1)},
	\end{gather}
	and \begin{eqnarray}
			\left( \frac{\partial}{\partial z^{\prime}} \right) ^a&=&\frac{g^{ab}\left( dz^{\prime} \right) _b}{g^{z^{\prime}z^{\prime}}}\nonumber\\
			&=&\frac{32l_ul_vz^4\left( 4u\left( uv+z^2 \right) -l_{u}^{2}v \right) F^{\prime}\left( z^{\prime} \right)}{A}\left( du \right) ^a
			+\frac{32l_ul_vz^4\left( 4v\left( uv+z^2 \right) -l_{v}^{2}u \right) F^{\prime}\left( z^{\prime} \right)}{A}\left( dv \right) ^a \nonumber\\
			&&+\frac{{ 4l_ul_vz^3\left( 16z^4-\left( \left( l_{u}^{2}-4u^2 \right) \left( l_{v}^{2}-4v^2 \right) \right) \right)}}{A}\left( dz \right) ^a,
		\end{eqnarray}
	where 
	\begin{eqnarray}
		A&=&\left( \left( l_u+2u \right) \left( l_v-2v \right) -4z^2 \right) \left( \left( l_u-2u \right) \left( l_v+2v \right) -4z^2 \right) \left( \left( l_u-2u \right) \left( l_v-2v \right) +4z^2 \right) \nonumber\\
		&&\times\left( \left( l_u+2u \right) \left( l_v+2v \right) +4z^2 \right) 
	\end{eqnarray}
	The transformations of the basis fields between $(u', v', z')$ and $(u,v,z)$ are as follows, 
	\begin{equation}
		\begin{aligned}
		\left( \frac{\partial}{\partial u^{\prime}} \right) ^a&=\frac{\partial u}{\partial u^{\prime}}\left( \frac{\partial}{\partial u} \right) ^a+\frac{\partial v}{\partial u^{\prime}}\left( \frac{\partial}{\partial v} \right) ^a+\frac{\partial z}{\partial u^{\prime}}\left( \frac{\partial}{\partial z} \right) ^a,\\
		\left( \frac{\partial}{\partial v^{\prime}} \right) ^a&=\frac{\partial u}{\partial v^{\prime}}\left( \frac{\partial}{\partial u} \right) ^a+\frac{\partial v}{\partial v^{\prime}}\left( \frac{\partial}{\partial v} \right) ^a+\frac{\partial z}{\partial v^{\prime}}\left( \frac{\partial}{\partial z} \right) ^a,\\
		\left( \frac{\partial}{\partial z^{\prime}} \right) ^a&=\frac{\partial u}{\partial z^{\prime}}\left( \frac{\partial}{\partial u} \right) ^a+\frac{\partial v}{\partial z^{\prime}}\left( \frac{\partial}{\partial v} \right) ^a+\frac{\partial z}{\partial z^{\prime}}\left( \frac{\partial}{\partial z} \right) ^a.	\label{73}
	\end{aligned}
	\end{equation}
They are connected by the Jacobian matrix and the inverse of the Jacobian matrix is
	\begin{equation}
		J^{-1}=\left( \begin{matrix}
			\frac{\partial u^{\prime}}{\partial u}&		\frac{\partial v^{\prime}}{\partial u}&		\frac{\partial z^{\prime}}{\partial u}\\
			\frac{\partial u^{\prime}}{\partial v}&		\frac{\partial v^{\prime}}{\partial v}&		\frac{\partial z^{\prime}}{\partial v}\\
			\frac{\partial u^{\prime}}{\partial z}&		\frac{\partial v^{\prime}}{\partial z}&		\frac{\partial z^{\prime}}{\partial z}\\
		\end{matrix} \right). 
	\end{equation}
The explicit forms of the entries can be readily get, therefore, we do not intend to present them explicitly because the expressions are rather complicated.
	Note that after integrating $
	\frac{\partial u^{\prime}}{\partial z}
	$ with respect to $z$, we get 
	\begin{equation}
		u^{\prime}\left( u,v,z \right) =\int{\frac{\partial u^{\prime}}{\partial z}dz}+C\left( u,v \right) ,
	\end{equation}
	 where $C(u,v)$ is a function of $u,v$. As $z\rightarrow 0$, the Rindler transformation in the bulk should be the same as the Rindler transformation on the boundary, from which we can determine $C(u,v)$ and thus obtain $u^\prime(u,v,z)$. $v^\prime(u,v,z)$ can be obtained in the same way, and the final results are
	  \begin{eqnarray}
	 	u^{\prime}=\frac{1}{4}\log \frac{l_{v}^{2}\left( l_u+2u \right) ^2-4\left( l_uv+2\left( uv+z^2 \right) \right) ^2}{l_{v}^{2}\left( l_u-2u \right) ^2-4\left( l_uv-2\left( uv+z^2 \right) \right) ^2}
	 			,\label{85}
	 		\\
	 		v^{\prime}=\frac{1}{4}\log \frac{l_{u}^{2}\left( l_v+2v \right) ^2-4\left( l_vu+2\left( uv+z^2 \right) \right) ^2}{l_{u}^{2}\left( l_v-2v \right) ^2-4\left( l_vu-2\left( uv+z^2 \right) \right) ^2}.\label{86}
	 \end{eqnarray}
	The three equations~\eqref{67},~\eqref{85}, and~\eqref{86} together constitute the Rindler transformation in the bulk. Under this transformation, the Rindler AdS metric  has the following form
	\begin{equation}
		ds^2=du^{\prime2}+dv^{\prime2}+2F\left( z^{\prime} \right) du^{\prime}dv^{\prime}+\frac{F^{\prime}\left( z^{\prime} \right) ^2}{4\left( F\left( z^{\prime} \right) ^2-1 \right)}dz^{\prime 2}.\label{87}
	\end{equation}
	It is easy to see that equations~\eqref{85} and~\eqref{86} are independent of the choice of $z^\prime$ (at least when  $g_{u^\prime z^\prime}=g_{v^\prime z^\prime}=0$), and the first two columns of matrix $J^{-1}$ are also independent of the choice of $z^\prime$.
	
	The radius of the horizon in Rindler AdS is
	\begin{equation}
		z^\prime_h=F^{-1}(1).\label{88}
	\end{equation}
	In order to obtain the thermal circle of Rindler AdS, we can expand the metric~\eqref{87} near the horizon~\cite{hartman2015lectures}. Let $z^\prime=z^\prime_h(1+\epsilon^2)$, as $\epsilon\rightarrow0$ we have
	\begin{equation}
		ds^2\approx \frac{1}{2}F^{-1}\left( 1 \right) F^{\prime}\left( F^{-1}\left( 1 \right) \right) d\epsilon ^2+\left( du^\prime+dv^\prime \right) ^2+2F^{-1}\left( 1 \right) F^{\prime}\left( F^{-1}\left( 1 \right) \right) \epsilon ^2du^{\prime}dv^{\prime}.
	\end{equation}
Now define a new coordinate $\phi=u^\prime+v^\prime$. If $\phi$ remains unchanged, that is, $d\phi=0$, then
	\begin{equation}
		\begin{aligned}
			ds^2&\approx \frac{1}{2}F^{-1}\left( 1 \right) F^{\prime}\left( F^{-1}\left( 1 \right) \right) d\epsilon ^2+\left( { du'+dv'} \right) ^2+2F^{-1}\left( 1 \right) F^{\prime}\left( F^{-1}\left( 1 \right) \right) \epsilon ^2du^{\prime}dv^{\prime}
			\\
			&=\frac{1}{2}F^{-1}\left( 1 \right) F^{\prime}\left( F^{-1}\left( 1 \right) \right) d\epsilon ^2-2F^{-1}\left( 1 \right) F^{\prime}\left( F^{-1}\left( 1 \right) \right) \epsilon ^2du^{\prime2}
			\\
			&=\frac{1}{2}F^{-1}\left( 1 \right) F^{\prime}\left( F^{-1}\left( 1 \right) \right) \left( d\epsilon ^2-\epsilon ^2\left( d\left( 2u^{\prime} \right) \right) ^2 \right) .
		\end{aligned}
	\end{equation}
Therefore,  the thermal circle is
\begin{equation}
	\left( u^{\prime},v^{\prime} \right) \sim \left( u^{\prime}+i\pi ,v^\prime-i\pi \right) ,
\end{equation}
which is consistent with the results in the field theory.

In addition, we can also calculate the Killing vector fields of Rindler AdS spacetime. For $F(z^\prime)>1$, the Killing vector fields are 
\begin{eqnarray}
		\xi _1&=&\partial _{u^{\prime}},\label{92}
		\\
		\xi _2&=&\partial _{v^{\prime}},
		\\
		\xi _3&=&-\frac{e^{2v^{\prime}}}{\sqrt{F\left( z^{\prime} \right) ^2-1}}\partial _{u^{\prime}}+\frac{F\left( z^{\prime} \right) e^{2v^{\prime}}}{\sqrt{F\left( z^{\prime} \right) ^2-1}}\partial _{v^{\prime}}-\frac{2\sqrt{F\left( z^{\prime} \right) ^2-1}e^{2v^{\prime}}}{F^{\prime}\left( z^{\prime} \right)}\partial _{z^{\prime}},
		\\
		\xi _4&=&-\frac{e^{-2v^{\prime}}}{\sqrt{F\left( z^{\prime} \right) ^2-1}}\partial _{u^{\prime}}+\frac{F\left( z^{\prime} \right) e^{-2v^{\prime}}}{\sqrt{F\left( z^{\prime} \right) ^2-1}}\partial _{v^{\prime}}+\frac{2\sqrt{F\left( z^{\prime} \right) ^2-1}e^{-2v^{\prime}}}{F^{\prime}\left( z^{\prime} \right)}\partial _{z^{\prime}},
		\\
		\xi _5&=&\frac{F\left( z^{\prime} \right) e^{2u^{\prime}}}{\sqrt{F\left( z^{\prime} \right) ^2-1}}\partial _{u^{\prime}}-\frac{e^{2u^{\prime}}}{\sqrt{F\left( z^{\prime} \right) ^2-1}}\partial _{v^{\prime}}-\frac{2e^{2u^{\prime}}\sqrt{F\left( z^{\prime} \right) ^2-1}}{F^{\prime}\left( z^{\prime} \right)}\partial _{z^{\prime}},
		\\
		\xi _6&=&\frac{F\left( z^{\prime} \right) e^{-2u^{\prime}}}{\sqrt{F\left( z^{\prime} \right) ^2-1}}\partial _{u^{\prime}}-\frac{e^{-2u^{\prime}}}{\sqrt{F\left( z^{\prime} \right) ^2-1}}\partial _{v^{\prime}}+\frac{2e^{-2u^{\prime}}\sqrt{F\left( z^{\prime} \right) ^2-1}}{F^{\prime}\left( z^{\prime} \right)}\partial _{z^{\prime}}.\label{97}
\end{eqnarray}

\subsection{ The thermal entropy of Rindler AdS}
 According to the Rindler method, the entanglement entropy in the field theory should be equal to the thermal entropy of Rindler AdS. Therefore, in this subsection, we will compute the black hole thermal entropy in TMG. The action contains the Einstein-Hilbert term, the cosmological constant term, and the Chern-Simons term 
	\begin{equation}
	\mathcal{S} _{\mathrm{TMG}}=\frac{1}{16\pi G}\int{d^3}x\sqrt{-g}\left[ R+2+\frac{1}{2\mu}\varepsilon ^{\alpha \beta \gamma}\left( {\Gamma ^{\rho}}_{\alpha \sigma}\partial _{\beta}{\Gamma ^{\sigma}}_{\alpha \rho}+\frac{2}{3}{\Gamma ^{\rho}}_{\alpha \sigma}{\Gamma ^{\sigma}}_{\beta \eta}{\Gamma ^{\eta}}_{\gamma \rho} \right) \right] .
	\end{equation}
where $\mu$ is a real coupling constant. The central charge of the dual CFT~\cite{Kraus:2005zm} is 
	\begin{equation}
		 c_L=\frac{3}{2G}\left(1+\frac{1}{\mu}\right),\qquad c_R=\frac{3}{2G}\left(1-\frac{1}{\mu}\right).
	\end{equation}
	We can directly calculate the thermal entropy, and the contribution of the Chern-Simons term to the thermal entropy~\cite{Jiang:2017ecm,Tachikawa:2006sz} is
	\begin{equation}
		S_{\text{CS}}=\frac{1}{4G\mu}\int_\Sigma\Gamma_N,\qquad\Gamma_N=-\frac{1}{2}{\epsilon^\mu}_\sigma{\Gamma^\sigma}_{\mu\rho}dx^\rho,
	\end{equation}
	where $\Sigma$ is the surface $z^\prime=z^\prime_h$, where $z^\prime_h$ is the horizon radius. The nonvanishing connection coefficients can be obtained from the metric~\eqref{87} as
\begin{eqnarray}
		{\Gamma ^{u^{\prime}}}_{u^{\prime}z^{\prime}}&=&{\Gamma ^{u^{\prime}}}_{z^{\prime}u^{\prime}}=\frac{F\left( z^{\prime} \right) F^{\prime}\left( z^{\prime} \right)}{2\left( F\left( z^{\prime} \right) ^2-1 \right)},\qquad {\Gamma ^{u^{\prime}}}_{v^{\prime}z^{\prime}}={\Gamma ^{u^{\prime}}}_{z^{\prime}v^{\prime}}=\frac{F^{\prime}\left( z^{\prime} \right)}{2-2F\left( z^{\prime} \right) ^2},\\
	{\Gamma ^{v^{\prime}}}_{u^{\prime}z^{\prime}}&=&{\Gamma ^{v^{\prime}}}_{z^{\prime}u^{\prime}}=\frac{F^{\prime}\left( z^{\prime} \right)}{2-2F\left( z^{\prime} \right) ^2},\qquad {\Gamma ^{v^{\prime}}}_{v^{\prime}z^{\prime}}={{\Gamma ^{v^{\prime}}}_{z^{\prime}v^{\prime}}}=\frac{F\left( z^{\prime} \right) F^{\prime}\left( z^{\prime} \right)}{2\left( F\left( z^{\prime} \right) ^2-1 \right)},\\
	{\Gamma ^{z^{\prime}}}_{u^{\prime}v^{\prime}}&=&{\Gamma ^{z^{\prime}}}_{v^{\prime}u^{\prime}}=\frac{2-2F\left( z^{\prime} \right) ^2}{F^{\prime}\left( z^{\prime} \right)},\qquad {\Gamma ^{z^{\prime}}}_{z^{\prime}z^{\prime}}=\frac{F\left( z^{\prime} \right) F^{\prime}\left( z^{\prime} \right)}{1-F\left( z^{\prime} \right) ^2}+\frac{F^{\prime\prime}\left( z^{\prime} \right)}{F^{\prime}\left( z^{\prime} \right)}.
\end{eqnarray}
From~\eqref{92}-\eqref{97}, we know that the Killing vector field orthogonal to $\Sigma$ is	
\begin{equation}
	\xi=\partial_{u^\prime}-\partial_{v^\prime}.
\end{equation}
The binormal vector $\epsilon_{ab}$ on $\Sigma$ satisfies 
\begin{equation}
	k\epsilon_{ab}|_\Sigma=\nabla_{a}\xi_b|_\Sigma,
\end{equation}
where $k$ is a constant. Since the binormal vectors should also satisfy 
\begin{equation}
	\epsilon_{ab}\epsilon^{ab}=-2,
\end{equation}
 we take $k$ as
 \begin{equation}
 	k = -\left. \sqrt{-\frac{1}{2}\nabla _a\xi _b\nabla ^a\xi ^b} \right|_{\varSigma}= -2.
 \end{equation}
Then we get
\begin{equation}
	S_{\text{CS}}=\frac{\Delta u^\prime-\Delta v^\prime}{4G\mu}.\label{109}
\end{equation}
For our regularized intervals~\eqref{57}, 
\begin{equation}
\Delta u^{\prime}=\frac{1}{2}\log \frac{l_{u}^{2}}{\varepsilon _{u1}\varepsilon _{u2}},\qquad \Delta v^{\prime}=\frac{1}{2}\log \frac{l_{v}^{2}}{\varepsilon _{v1}\varepsilon _{v2}}.\label{110}
\end{equation}
The total thermal entropy of Rindler AdS spacetime is
\begin{equation}
	\begin{aligned}
		S_{\mathrm{Rindler}}&=S_{BH}+S_{CS}\\
		&=\frac{\Delta u^{\prime}+\Delta v^{\prime}}{4G}+\frac{\Delta u^{\prime}-\Delta v^{\prime}}{4G\mu},
	\end{aligned}
\end{equation}
where $S_{BH}$ is the Bekenstein-Hawking entropy. Notice that 
\begin{equation}
	\frac{1}{4G}=\frac{c_L+c_R}{12},\qquad \frac{1}{4G\mu}=\frac{c_L-c_R}{12}
\end{equation}
and 
\begin{equation}
	\Delta u^{\prime}+\Delta v^{\prime}=\frac{1}{2}\log \frac{l_{u}^{2}l_{v}^{2}}{\varepsilon _{u1}\varepsilon _{u2}\varepsilon _{v1}\varepsilon _{v2}},\qquad \Delta u^{\prime}-\Delta v^{\prime}=\frac{1}{2}\log \frac{l_{u}^{2}\varepsilon _{v1}\varepsilon _{v2}}{l_{v}^{2}\varepsilon _{u1}\varepsilon _{u2}},
\end{equation}
 then we get,
 \begin{equation}
 	S_{\mathrm{Rindler}}=\frac{c_L+c_R}{24}\log \frac{l_{u}^{2}l_{v}^{2}}{\varepsilon _{u1}\varepsilon _{v1}\varepsilon _{u2}\varepsilon _{v2}}+\frac{c_L-c_R}{24}\log \frac{l_{u}^{2}\varepsilon _{v1}\varepsilon _{v2}}{l_{v}^{2}\varepsilon _{u1}\varepsilon _{u2}}.\label{srindler}
 \end{equation}
  Therefore, the entropy of Rindler AdS spacetime is consistent with the calculation of entanglement entropy in field theory~\eqref{3.49}. Moreover, the contribution of the Chern-Simons term to the black hole thermal entropy comes from the last term in the equation \eqref{srindler}
 \begin{equation}
 	S_{CS}=\frac{c_L-c_R}{24}\log \frac{l_{u}^{2}\varepsilon _{v1}\varepsilon _{v2}}{l_{v}^{2}\varepsilon _{u1}\varepsilon _{u2}},
 \end{equation}
  which is exactly the anomaly contribution in the field theory.
\section{Entanglement entropy of zero temperature CFT on a cylinder and finite temperature CFT in a plane}\label{s5}
 It is well-known that the constraint equation \eqref{surface} can be reformulated by the following coordinates ~\cite{Aharony:1999ti,Hartman:2013qma} 
\begin{equation}
\begin{cases}
	y_0=\cosh \rho \cos \tau,\\
	y_1=\cosh \rho \sin \tau,\\
	y_2=\sinh \rho \sin \theta,\\
	y_3=-\sinh \rho \cos \theta,\\
\end{cases}
\end{equation}
 Here we have set the radius of AdS to be $L=1$. Therefore, the metric of AdS becomes
 \begin{equation}
 	ds^2=-\cosh^2\rho d\tau^2+d\rho^2+\sinh\rho^2 d\theta^2.
 \end{equation}
Its boundary is a cylinder with $\rho=\infty$. On the other hand, according to Eqs.~\eqref{14} and~\eqref{15}, we know that to write AdS$_3$ in Poincar\'e coordinates, we have
\begin{equation}
	\begin{cases}
		y_0=\frac{1-t^2+x^2+z^2}{2z},\\
		y_1=\frac{t}{z},\\
		y_2=\frac{x}{z},\\
		y_3=\frac{-1-t^2+x^2+z^2}{2z}.
	\end{cases}
\end{equation}
Therefore, the relationship between the global coordinates and the Poincar\'e coordinates is
\begin{equation}
	\left\{ \begin{array}{cc}
		t&=\frac{\cosh \rho \sin \tau}{\cos \tau \cosh \rho +\cos \theta \sinh \rho},\\
		x&=\frac{\sin \theta \sinh \rho}{\cosh \rho \cos \tau +\sinh \rho \cos \theta},\\
		z&=\frac{1}{\cosh \rho \cos \tau +\sinh \rho \cos \theta}.
	\end{array} \right. 
\end{equation}
On the boundary $\rho\to\infty$, this is precisely a conformal transformation, 
\begin{equation}
\left\{ \begin{array}{c}
	t=\frac{\sin \tau}{\cos \tau+\cos \theta},\\
	x=\frac{\sin \theta}{\cos \tau+\cos \theta}.\\
\end{array} \right.
\end{equation}
One can rewrite it in the form of light-like coordinates as
\begin{equation}
	\left\{ \begin{array}{c}
		u=\tan \frac{U}{2},\\
		v=\tan \frac{V}{2}.\label{117}
	\end{array} \right. 
\end{equation}
where $U=\theta+\tau$, $V=\theta-\tau$, are exactly the transformations between the plane and the Lorentzian cylinder in two-dimensional Minkowski spacetime~\cite{Mack:1988nf}. In principle, \eqref{117} together with \eqref{31} and \eqref{32} can give a Rindler transformation that maps a zero-temperature CFT on a cylinder to a finite-temperature CFT on a plane. Consider an interval on the cylinder (the upper label of $\mathcal{I}^c$ represents cylinder)
\begin{equation}
	\mathcal{I} ^c\equiv \left\{ \left( U,V \right) \bigg|U=l_U\left( s-\frac{1}{2} \right) ,V=l_V\left( s-\frac{1}{2} \right) ,s\in \left[ 0,1 \right] \right\} 
\end{equation}
 Under the transformations, it becomes an interval on the plane,
 \begin{equation}
 	\mathcal{I} ^p\equiv \left\{ \left( u,v \right) \bigg|u=\tan \left[ \frac{l_U}{2}\left( s-\frac{1}{2} \right) \right] ,v=\tan \left[ \frac{l_V}{2}\left( s-\frac{1}{2} \right) \right] ,s\in \left[ 0,1 \right] \right\} 
 \end{equation}
 Thus, computing the entanglement entropy of $\mathcal I^p$ on the plane can obtain the entanglement entropy of the interval $\mathcal{I} ^c$. Considering the entanglement entropy formula~\eqref{76}, as well as the length of the interval \footnote{The length of the interval here refers to the length of its decomposition into two quasi-light directions. For example, for interval $ \mathcal{I}$~\eqref{22}, the length of the interval refers to $l_u$ and $l_v$.} and the UV cutoff, we can obtain the entanglement entropy. The length of interval $ \mathcal{I}^p$ are
  \begin{equation}
  	l_{u}^{p}=2\tan \frac{l_U}{4},\qquad l_{v}^{p}=2\tan \frac{l_V}{4}.
  \end{equation}
If we assume that the regulated interval on the cylinder is
\begin{equation}
	\mathcal{I} ^{\prime c}\equiv \left\{ \left( U,V \right) \bigg|U=l_U\left( s-\frac{1}{2} \right) ,V=l_V\left( s-\frac{1}{2} \right) ,s\in \left[ \varepsilon _1,1-\varepsilon _2 \right] \right\} 
\end{equation}
and let
\begin{gather}
	\varepsilon _{U1}=l_U\varepsilon _1,\qquad \varepsilon _{U2}=l_U\varepsilon _2,\label{3.95}
	\\
	\varepsilon _{V1}=l_V\varepsilon _1,\qquad \varepsilon _{V2}=l_V\varepsilon _2, \label{3.96}
\end{gather}
the regulated interval can be readily obtained after the conformal transformation~\eqref{117},
\begin{equation}
	{\mathcal{I} ^{\prime}}^p\equiv \left\{ \left( u,v \right) \bigg|u=\tan \left[ \frac{l_U}{2}\left( s-\frac{1}{2} \right) \right] ,v=\tan \left[ \frac{l_V}{2}\left( s-\frac{1}{2} \right) \right] ,s\in \left[ \varepsilon _1,1-\varepsilon _2 \right] \right\} 
\end{equation}
 The cutoffs can be read from the regulated interval $\mathcal{I}^{\prime p}$ as
\begin{gather}
	\varepsilon _{u1}^{p}\equiv \tan \left( \frac{1}{2}\varepsilon _{U1}-\frac{l_U}{4} \right) -\tan \left( -\frac{l_U}{4} \right) ,\qquad \varepsilon _{u2}^{p}\equiv \tan \frac{l_U}{4}-\tan \left( \frac{l_U}{4}-\frac{1}{2}\varepsilon _{U2} \right) ,
	\\
	\varepsilon _{v1}^{p}\equiv \tan \left( \frac{1}{2}\varepsilon _{V1}-\frac{l_V}{4} \right) -\tan \left( -\frac{l_V}{4} \right) ,\qquad \varepsilon _{v2}^{p}\equiv \tan \frac{l_V}{4}-\tan \left( \frac{l_V}{4}-\frac{1}{2}\varepsilon _{V2} \right) .
\end{gather}
Substituting them into Eq.~\eqref{76}, one can obtain the entanglement entropy on the cylinder
\begin{equation}
	\begin{aligned}
		S&=\frac{c_L}{12}\log \frac{\left( l_{u}^{p} \right) ^2}{\varepsilon _{u1}^{p}\varepsilon _{u2}^{p}}+\frac{c_R}{12}\log \frac{\left( l_{v}^{p} \right) ^2}{\varepsilon _{v1}^{p}\varepsilon _{v2}^{p}}\\
		&=\frac{c_L}{12}\log \frac{\left( 2\tan \frac{l_U}{4} \right) ^2}{\left( \tan \left( \frac{1}{2}\varepsilon _{U1}-\frac{l_U}{4} \right) -\tan \left( -\frac{l_U}{4} \right) \right) \left( \tan \frac{l_U}{4}-\tan \left( \frac{l_U}{4}-\frac{1}{2}\varepsilon _{U2} \right) \right)}\\
		&+\frac{c_R}{12}\log \frac{\left( 2\tan \frac{l_V}{4} \right) ^2}{\left( \tan \left( \frac{1}{2}\varepsilon _{V1}-\frac{l_V}{4} \right) -\tan \left( -\frac{l_V}{4} \right) \right) \left( \tan \frac{l_V}{4}-\tan \left( \frac{l_V}{4}-\frac{1}{2}\varepsilon _{V2} \right) \right)}\\
		&\approx \frac{c_L}{12}\log \frac{4\sin ^2\frac{l_U}{2}}{\varepsilon _{U1}\varepsilon _{U2}}+\frac{c_R}{12}\frac{4\sin ^2\frac{l_V}{2}}{\varepsilon _{V1}\varepsilon _{V2}}\\
		&
		=\frac{c_L+c_R}{24}\log \frac{16\sin ^2\frac{l_U}{2}\sin ^2\frac{l_V}{2}}{\varepsilon _{U1}\varepsilon _{V1}\varepsilon _{U2}\varepsilon _{V2}}+\frac{c_L-c_R}{24}\log \frac{\sin ^2\frac{l_U}{2}\varepsilon _{V1}\varepsilon _{V2}}{\sin ^2\frac{l_V}{2}\varepsilon _{U1}\varepsilon _{U2}},
			\label{127}
	\end{aligned}
\end{equation}
in which the last term is the contribution of the Chern-Simons term to the black hole entropy in the bulk
\begin{equation}
S_{CS}=
\frac{c_L-c_R}{24}\log \frac{\sin ^2\frac{l_U}{2}\varepsilon _{V1}\varepsilon _{V2}}{\sin ^2\frac{l_V}{2}\varepsilon _{U1}\varepsilon _{U2}}.\label{3.122}
\end{equation}
As in the previous discussion, these cutoffs do not generally need to satisfy equations \eqref{3.95} and \eqref{3.96}. It is worth noting that the results we have obtained are for CFT with a spatial circle
\begin{equation}
	U\sim U+2\pi ,\qquad V\sim V+2\pi .
\end{equation}
If we consider a CFT with an arbitrary spatial circle
\begin{equation}
	U\sim U+L_U,\qquad V\sim V+L_V,
\end{equation}
 we need to make the following substitution in~\eqref{127} and~\eqref{3.122},
 \begin{gather}
l_U\rightarrow \frac{l_U}{L_U/2\pi},\qquad l_V\rightarrow \frac{l_V}{L_V/2\pi},
\\
\varepsilon _{U1}\rightarrow \frac{\varepsilon _{U1}}{L_U/2\pi},\qquad \varepsilon _{V1}\rightarrow \frac{\varepsilon _{V1}}{L_V/2\pi},
\\
\varepsilon _{U2}\rightarrow \frac{\varepsilon _{U2}}{L_U/2\pi},\qquad \varepsilon _{V2}\rightarrow \frac{\varepsilon _{V2}}{L_V/2\pi}.
 \end{gather}
 Ultimately, we obtain
\begin{gather}
	S=\frac{c_L+c_R}{24}\log \frac{L_{U}^{2}L_{V}^{2}\sin ^2\frac{\pi l_U}{L_U}\sin ^2\frac{\pi l_V}{L_V}}{\pi ^4\varepsilon _{U1}\varepsilon _{V1}\varepsilon _{U2}\varepsilon _{V2}}+\frac{c_L-c_R}{24}\log \frac{L_{U}^{2}\sin ^2\frac{\pi l_U}{L_U}\varepsilon _{V1}\varepsilon _{V2}}{L_{V}^{2}\sin ^2\frac{\pi l_V}{L_V}\varepsilon _{U1}\varepsilon _{U2}},\label{3.118}
\end{gather}
and the contribution from the Chern-Simons term is
\begin{gather}
	S_{CS}=\frac{c_L-c_R}{24}\log \frac{L_{U}^{2}\sin ^2\frac{\pi l_U}{L_U}\varepsilon _{V1}\varepsilon _{V2}}{L_{V}^{2}\sin ^2\frac{\pi l_V}{L_V}\varepsilon _{U1}\varepsilon _{U2}}.
\end{gather}
Based on these formulae, replacing the spatial circle with a thermal circle ($L_U\rightarrow i\beta_U,L_V\rightarrow i\beta_V$) will directly yield the entanglement entropy of the finite-temperature CFT and the corrections of the Chern-Simons term to the thermal entropy in the bulk. They are,
\begin{eqnarray}
S&=&\frac{c_L+c_R}{24}\log \frac{\beta _{U}^{2}\beta _{V}^{2}\sinh ^2\frac{\pi l_U}{\beta _U}\sinh ^2\frac{\pi l_V}{\beta _V}}{\pi ^4\varepsilon _{U1}\varepsilon _{V1}\varepsilon _{U2}\varepsilon _{V2}}+\frac{c_L-c_R}{24}\log \frac{\beta _{U}^{2}\sinh ^2\frac{\pi l_U}{\beta _U}\varepsilon _{V1}\varepsilon _{V2}}{\beta _{V}^{2}\sinh ^2\frac{\pi l_V}{\beta _V}\varepsilon _{U1}\varepsilon _{U2}},\label{3.109}
\\
S_{CS}&=&\frac{c_L-c_R}{24}\log \frac{\beta _{U}^{2}\sinh ^2\frac{\pi l_U}{\beta _U}\varepsilon _{V1}\varepsilon _{V2}}{\beta _{V}^{2}\sinh ^2\frac{\pi l_V}{\beta _V}\varepsilon _{U1}\varepsilon _{U2}}.
\end{eqnarray}
It is obvious that if we take $c_L=c_R$ and choose the subsystem to be an interval on the equal-time surface, our previous results can all return to the results of the two-dimensional conformal field theory \cite{Calabrese:2004eu,Calabrese:2009qy}. For $c_L\neq c_R$, equations \eqref{3.118} and \eqref{3.109} do not necessarily have anomalous contributions, for example, there are no anomalous contributions when \begin{equation}
\begin{array}{c}
	L_{U}^{2}\sin ^2\frac{\pi l_U}{L_U}\varepsilon _{V1}\varepsilon _{V2}=L_{V}^{2}\sin ^2\frac{\pi l_V}{L_V}\varepsilon _{U1}\varepsilon _{U2},\\
	\beta _{U}^{2}\sinh ^2\frac{\pi l_U}{\beta _U}\varepsilon _{V1}\varepsilon _{V2}=\beta _{V}^{2}\sinh ^2\frac{\pi l_V}{\beta _V}\varepsilon _{U1}\varepsilon _{U2.}
\end{array}
\end{equation}

\section{Conclusions}\label{s6}
In this paper, we studied the entanglement entropy of the two-dimensional CFT with gravitational anomalies and derived some of their important properties. By reviewing AdS$_3$ spacetime and calculating Killing vector fields, we derived the global conformal transformation generators of the boundary CFT. Using the generalized Rindler method, we derived the Rindler transformation in two-dimensional planar CFT and calculated the entanglement entropy of CFT with gravitational anomalies in the field theory. In addition, we also derived the Rindler transformation in the bulk and found that the transformation of the $u,v$ coordinates in the bulk has nothing to do with the transformation of $z$ coordinate. From the relationship between global coordinates and Poincar\'e coordinates in the bulk, we derived the transformation between the plane and Lorentzian cylinder on the boundary and found that this transformation is consistent with that in Mack's paper~\cite{Mack:1988nf}. Finally, using conformal transformations, we calculated the entanglement entropy of CFT on a plane and derived the entanglement entropy formula for zero-temperature CFT on a cylinder and the entanglement entropy of CFT with temperature on a plane.  We found that the entanglement entropy we calculated in field theory is indeed equal to the black hole entropy in Rindler AdS. Specifically, for a zero-temperature CFT on a plane, if the total boost angle ($\delta\kappa_1+\delta\kappa_2$) of the cutoffs relative to the space-like diagonal of the DOD of the subregion is not zero, then there will be an anomalous contribution to the entanglement entropy. These results are important for further understandings of the two-dimensional CFT with gravitational anomalies and may deepen our understanding of holographic entanglement entropy.

\appendix
\section{Derive the Rindler transformation}\label{A}
	Considering only $	u^\prime=f(u)$, we can obtain from equations \eqref{19} and \eqref{21}:
\begin{equation}
	\begin{aligned}
		\partial _{u^{\prime}}&=\frac{du}{du^{\prime}}\partial _u
		=a_0l_0+a_+l_++a_-l_-
		=\left( a_0u-a_+-u^2a_- \right) \partial _u.
	\end{aligned}\label{24}
\end{equation}
This implies that we only need to solve 
\begin{equation}
	\frac{du^{\prime}}{du}=\frac{1}{a_0u-a_+-u^2a_-}
\end{equation}
to obtain the Rindler transformation up to some undetermined constants $c$. One can easily get
\begin{equation}
	u^\prime=-\frac{2\arctan\frac{2a_-u-a_0}{\sqrt{4a_+a_--a^2_0}}}{\sqrt{4a_+a_--a^2_0}}+c
\end{equation}
and then solve for $u$ as 
\begin{equation}
	u=\frac{a_0-\sqrt{4a_+a_--a_{0}^{2}}}{2a_-}\tan \left( \frac{1}{2}\sqrt{4a_+a_--a_{0}^{2}}u^{\prime} \right) .
\end{equation}
Here, without loss of generality, we have taken the constant $c$ to be $0$. It is easy to see that an imaginary periodicty exists in the $u'$ direction only when $4a_+a_--a^2_0<0$ is satisfied, and the imaginary periodicty is:
\begin{equation}
	\beta_{u^\prime}=\frac{2\pi}{\sqrt{a^2_0-4a_+a_-}}.\label{28}
\end{equation}
The generator of the modular flow is given by $k_t = \beta_{u'}(a_0u - a_+ - u^2a_-)\partial_u + \beta_{v'}\frac{dv}{dv'}\partial_v$. To ensure that the boundary of $\mathcal{I}$ remains invariant under $k_t$, we must have
\begin{equation}
	\begin{aligned}
		a_0\frac{l_u}{2}-a_+-(l_u/2)^2a_-&=0\\
		-a_0\frac{l_u}{2}-a_+-(-l_u/2)^2a_-&=0\label{29}
	\end{aligned}
\end{equation}
From \eqref{28} and \eqref{29}, we can obtain the following solution
\begin{equation}
	\left\{ \begin{array}{ccc}
		a_0&=&0,\\
		a_+&=&-\frac{l_u\pi}{2\beta _{u^{\prime}}},\\
		a_-&=&\frac{2\pi}{\beta _{u^{\prime}}l_u}.\\
	\end{array} \right. 
\end{equation}
Therefore, the Rindler transformation can be determined as follows
\begin{equation}
	u^\prime=\frac{\beta_{u^\prime}}{\pi}\arctan\!\text{h}\frac{2u}{l_u}.
\end{equation}
Similarly, one can get
\begin{equation}
	\begin{aligned}
		\partial _{v^{\prime}}&=\frac{dv}{dv^{\prime}}\partial _v=b_0\bar{l}_0+b_+\bar{l}_++b_-\bar{l}_-
		=\left( -b_0v+b_+v^2+b_- \right) \partial _v,
	\end{aligned}\label{33}
\end{equation}
By comparing equations~\eqref{24} and~\eqref{33}, we can take
\begin{equation}
	v^\prime=-\frac{\beta_{v^\prime}}{\pi}\arctan\!\text{h}\frac{2v}{l_v}
\end{equation}
with the periodicity
\begin{equation}
	v^\prime\sim v^\prime+i\beta_{v^\prime},\qquad \beta_{v^\prime}=\frac{2\pi}{b^2_0-4b_+b_-}.
\end{equation}
The constants in equation~\eqref{33} can be determined,
\begin{equation}
	\left\{ \begin{array}{ccc}
		b_0&=&0,\\
		b_+&=&\frac{2\pi}{\beta _{v^{\prime}}l_v},\\
		b_-&=&-\frac{l_v\pi}{2\beta _{v^{\prime}}}.\\
	\end{array} \right. 
\end{equation}
Since $\beta_{u^\prime}$ and $\beta_{v^\prime}$ do not appear in the expression of entangled entropy, the choice of $\beta_{u^\prime}$ and $\beta_{v^\prime}$ is arbitrary.

\acknowledgments

We thank Ren Jie for helpful discussions the Killing vectors. This work was partially supported by the National Natural Science Foundation of China (Grants No.11875095 and 12175008).





	
\end{document}